 \documentclass[10pt, final]{IEEEtran}

\usepackage{amsthm}

\usepackage{amssymb}
\usepackage{amsmath}
\usepackage{bbm}
\usepackage{mathrsfs}
\usepackage{cite}
\usepackage{bm,epsfig,amsthm,url}
\usepackage{indentfirst}
\usepackage{amsfonts}
\usepackage{epstopdf}
\usepackage{cases}
\usepackage[caption=false,font=scriptsize]{subfig}
\usepackage{xcolor}
\usepackage{mathtools}

\makeatletter
\renewcommand{\maketag@@@}[1]{\hbox{\m@th\normalsize\normalfont#1}}%
\makeatother
\usepackage{hyperref}
\usepackage{stfloats}

\newtheoremstyle{mystyle}{}{}{}{}{}{: }{0pt}{\indent \it{\thmname{#1}\thmnumber{ #2}\thmnote{#3}}}
\theoremstyle{mystyle}

\newtheorem{Proposition}{Proposition}

\usepackage[noend]{algpseudocode}
\usepackage{float}
\usepackage{algorithmicx}
\usepackage[linesnumbered,ruled,vlined]{algorithm2e}

\begin{document}

\title{\fontsize{20}{26}\selectfont Cooperative Rotatable IRSs for Wireless Communications: Joint Beamforming and Orientation Optimization}

\author{{Qiaoyan~Peng,~Qingqing~Wu,~Guangji~Chen,~Wen~Chen,~Shanpu~Shen,~Shaodan~Ma}
	\thanks{Q. Peng is with the Department of Electronic Engineering, Shanghai Jiao Tong University, Shanghai 200240, China, and also with the State Key Laboratory of Internet of Things for Smart City, University of Macau, Macao 999078, China (email: qiaoyan.peng@connect.um.edu.mo). 
	Q. Wu and W. Chen are with the Department of Electronic Engineering, Shanghai Jiao Tong University, Shanghai 200240, China (email: qingqingwu@sjtu.edu.cn; wenchen@sjtu.edu.cn).
	G. Chen is with Nanjing University of Science and Technology, Nanjing 210094, China (email: guangjichen@njust.edu.cn).
	S. Ma and S. Shen are with the State Key Laboratory of Internet of Things for Smart City, University of Macau, Macao 999078, China (email: shaodanma@um.edu.mo; shanpushen@um.edu.mo).
	}
}

\maketitle

\begin{abstract}
Rotatable intelligent reflecting surfaces (IRSs) introduce a new degree of freedom (DoF) for shaping wireless propagation by adaptively adjusting the orientation of IRSs. This paper considers an angle-dependent reflection model in a wireless communication system aided by two rotatable IRSs. Specifically, we study the joint design of the base station transmit beamforming, as well as the cooperative passive beamforming and orientation of the two IRSs, to maximize the received signal-to-noise ratio (SNR). Under the light-of-sight (LoS) channels, we first develop a particle swarm optimization (PSO) based method to determine the IRS rotation and derive an optimal rotation in a closed-form expression for a two-dimensional IRS deployment. Then, we extend the design to the general Rician fading channels by proposing an efficient alternating optimization and PSO (AO-PSO) algorithm. Numerical results validate the substantial gains achieved by the IRS rotation over fixed-IRS schemes and also demonstrate the superior performance of the double rotatable IRSs over a single rotatable IRS given a sufficient total number of IRS elements.
\end{abstract}

\begin{IEEEkeywords}
Alternating optimization (AO), intelligent reflecting surface (IRS), particle swarm optimization (PSO), rotatable IRS, rotation optimization.
\end{IEEEkeywords}

\section{Introduction}
Intelligent reflecting surfaces (IRSs) have recently been deemed as a cost-effective and energy-efficient technology for reshaping wireless propagation environments. The fundamental squared-power gain of a single passive IRS, i.e., scaling on the order of $\mathcal{O}(N^{2})$, with respect to (w.r.t.) the total number of reflecting elements $N$, achieved through passive beamforming has been established in \cite{wu_6G, chen_iot}. This highlights the potential of IRS-assisted transmission with a strong reflected path between the base station (BS) and the user. However, the performance of single-IRS-involved links is constrained by blockage, shadowing, and geometric placement. To overcome these limitations, communication systems aided by two or more cooperative IRSs have been investigated \cite{mei}. Proper coordination of the passive beamforming vectors at the two IRSs can significantly enhance the overall reflection gain, which increases on the order of $\mathcal{O}(N^{4})$ \cite{doubleIRS_zheng}. This higher-order cooperative beamforming gain highlights the significant potential of double IRSs for improving link reliability and system performance. 

Existing studies typically rely on idealized reflection models that are isotropic and orientation-independent. However, this assumption may lead to a mismatch between the IRS orientation and the incident/reflected signal directions, resulting in severe reflection loss and substantially degrading the achievable beamforming gain. Consequently, the corresponding results may not be applicable in practical deployment scenarios. To address this practical limitation and fully unleash the potential of IRS-assisted systems, the joint optimization of the passive beamforming and the deployment orientation has attracted considerable attention \cite{magazine,rotate1,rotate2}. A rotatable IRS has been proposed to further exploit spatial degrees of freedom (DoFs) by adaptively controlling the orientation of IRSs \cite{peng2025rotatable}. Specifically, the IRS can be mounted on a motor-driven shaft, enabling rotation control without changing the element positions. By mechanically adjusting the rotation of the IRS, the incident and reflected directions can be dynamically reshaped to achieve effective reflection. Previous studies have unveiled that proper orientation design of multiple IRSs can significantly improve the channel condition \cite{rotate21,rotate22,rotate23,rotate24}. Although the work \cite{rotate24} derived the sub-optimal IRS rotation angles in a closed-form expression, these results are limited to one-dimensional (1D) orientation adjustment. In contrast, two-dimensional (2D) orientation optimization provides additional spatial DoF, which enable more effective alignment between the incident and reflected signals.

Motivated by the above considerations, we study the fundamental design and performance of a cooperative double rotatable IRS-aided wireless communication system. Our goal is to maximize the received signal-to-noise ratio (SNR) at the user by jointly optimizing the BS transmit beamforming, the IRS passive beamforming, and the IRS rotation. Under the LoS channels, we develop a particle swarm optimization (PSO) based method to optimize the IRS rotations and derive the optimal rotation angles in a closed-form expression for a 2D IRS deployment. To solve the challenging non-convex optimization problem under Rician fading channels, an alternating optimization and PSO (AO-PSO) algorithm is proposed to obtain a high-quality sub-optimal solution. Numerical results demonstrate that IRS rotation brings notable gains over fixed-IRS schemes and further reveal that the double rotatable IRS outperforms its single-IRS counterpart when the total number of IRS elements is sufficiently large.

\section{System Model and Problem Formulation}
\label{System Model}
\begin{figure}[t]
	\centering
	\includegraphics[width=0.33\textwidth]{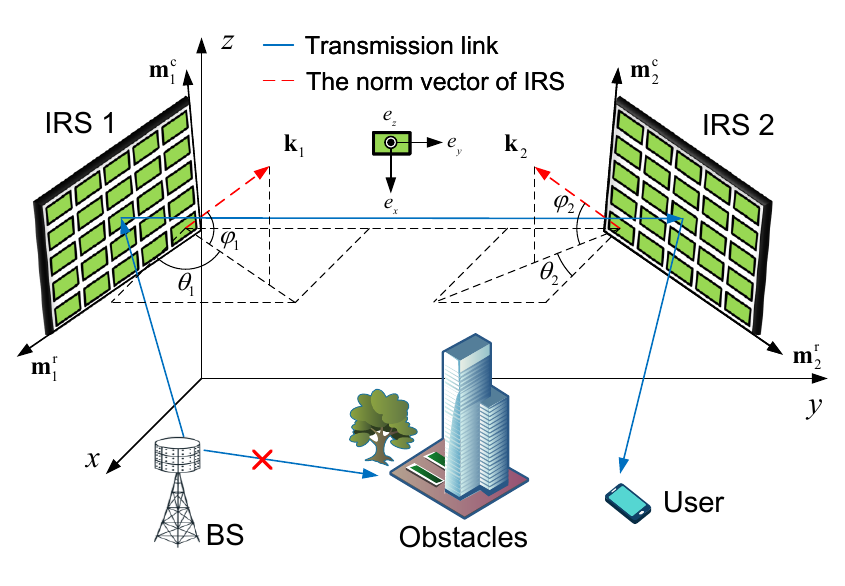}
	\caption{A wireless communication system enhanced by cooperative double rotatable IRSs.}
	\label{fig:systemmodel}
\end{figure}

As shown in Fig. \ref{fig:systemmodel}, we consider a cooperative double rotatable IRS-aided downlink communication system. We assume that a multi-antenna BS serves a single-antenna user through the cascaded BS-IRS 1-IRS 2-user reflection path, while the direct and single-reflection links are ignored in the heavy-blockage scenarios. The IRS 1 (IRS 2) consists of a total of $N_1 = N_1^{\mathrm{c}} \times N_1^{\mathrm{r}} (N_2 = N_2^{\mathrm{c}} \times N_2^{\mathrm{r}})$ elements, arranged in $N_1^{\mathrm{r}} (N_2^{\mathrm{r}})$ rows and $N_1^{\mathrm{c}} (N_2^{\mathrm{c}})$ columns. Considering a three-dimensional (3D) Cartesian coordinate system (CCS), the user is placed at $\mathbf{p}_\mathrm{U} = (x_\mathrm{U}, y_\mathrm{U}, 0)^T$.

The orientation matrices of the IRS 1 and IRS 2 are given by $\mathbf{\Omega}_1 = (\theta_1, \varphi_1)^T$ and $\mathbf{\Omega}_2 = (\theta_2, \varphi_2)^T$, respectively, where $\theta _1$ and $\varphi _1$ ($\theta _2$ and $\varphi _2$) denote the azimuth and elevation angles of the normal vector of the IRS 1 ${{\mathbf{k}}_1}$ (IRS 2 ${{\mathbf{k}}_2}$). Then, we have ${{\mathbf{k}}_1} = ( { \cos {\theta _1}\cos {\varphi _1},\sin {\theta _1}\cos {\varphi _1}, \sin {\varphi _1}} )^T$ and ${{\mathbf{k}}_2} = ( {\cos {\theta _2}\cos {\varphi _2}, - \sin {\theta _2}\cos {\varphi _2},\sin {\varphi _2}} )^T$. The reflecting elements of IRS 1 (IRS 2) are arranged along two orthogonal directions, denoted by ${\mathbf{m}}_1^{\mathrm{r}} \!=\! ( \sin {\theta _1}, - \cos {\theta _1},0 )^T$ and ${\mathbf{m}}_1^{\mathrm{c}} = (  - \cos {\theta _1}\sin {\varphi _1}, - \sin {\theta _1}\sin {\varphi _1},\cos {\varphi _1} )^T$ (${\mathbf{m}}_2^{\mathrm{c}} = (  - \cos {\theta _2}\sin {\varphi _2},\sin {\theta _2}\sin {\varphi _2},\cos {\varphi _2} )^T$ and ${\mathbf{m}}_2^{\mathrm{r}} = ( \sin {\theta _2},\cos {\theta _2},0 )^T$).

%

We map the reflecting element located at row $\bar N_{n_1}^\mathrm{r}$ and column $\bar N_{n_1}^\mathrm{c}$ as the element $n_1$ of IRS 1 with $\bar N_{n_1}^\mathrm{r} \in \mathcal{N}_1^\mathrm{r} = \left\{0,\cdots,N_{1}^\mathrm{r}-1\right\}$, $\bar N_{n_1}^\mathrm{c} \in \mathcal{N}_1^\mathrm{c} = \left\{0,\cdots,N_{1}^\mathrm{c}-1\right\}$, and $n_1 \in \mathcal{N}_1 = \left\{1,\cdots,N_1 \right\}$. Then, we have $\bar N_{{n_1}}^{\mathrm{r}} = \left\lfloor {\left( {{n_1} - 1} \right)/N_1^{\mathrm{c}}} \right\rfloor$ and $\bar N_{{n_1}}^{\mathrm{c}} = {n_1} - N_1^{\mathrm{c}}\bar N_{{n_1}}^{\mathrm{r}} - 1$. Similarly, the row and column indices of the $n_2$-th element of IRS 2 are given by $\tilde N_{{n_2}}^{\mathrm{r}} = \left\lfloor {\left( {{n_2} - 1} \right)/N_2^{\mathrm{c}}} \right\rfloor$, and $\tilde N_{{n_2}}^{\mathrm{c}} = {n_2} - N_2^{\mathrm{c}}\tilde N_{{n_2}}^{\mathrm{r}} - 1$ with $\tilde N_{n_2}^\mathrm{r} \in \mathcal{N}_2^\mathrm{r} = \left\{0,\cdots,N_{2}^\mathrm{r}-1\right\}$, $\tilde N_{n_2}^\mathrm{c} \in \mathcal{N}_2^\mathrm{c} = \left\{0,\cdots,N_{2}^\mathrm{c}-1\right\}$, and $n_2 \in \mathcal{N}_2 = \left\{1,\cdots,N_2 \right\}$. We denote the coordinates of the first elements of IRS 1 and IRS 2 as $\mathbf{p}_{1,1} = {\left( {x_{1,1},y_{1,1},z_{1,1}} \right)^T}$ and ${{\mathbf{p}}_{2,1}} = {\left( {x_{2,1},y_{2,1},z_{2,1}} \right)^T}$, respectively. Accordingly, the coordinates of the element $n_1$ of the IRS 1 and the element $n_2$ of the IRS 2 are given by ${{\mathbf{p}}_{1,{n_1}}} = {{\mathbf{p}}_{1,1}} + \bar N_{{n_1}}^{\mathrm{c}}l{\mathbf{m}}_1^{\mathrm{r}} + \bar N_{{n_1}}^{\mathrm{r}}l{\mathbf{m}}_1^{\mathrm{c}}, n_1 \in \mathcal{N}_1$ and ${{\mathbf{p}}_{2,{n_2}}} = {{\mathbf{p}}_{2,1}} + \tilde N_{{n_2}}^{\mathrm{c}}l{\mathbf{m}}_2^{\mathrm{r}} + \tilde N_{{n_2}}^{\mathrm{r}}l{\mathbf{m}}_2^{\mathrm{c}},	n_2 \in \mathcal{N}_2$, respectively,
where $l$ denotes the inter-element spacing of each IRS. 

For the transmit antenna array at the BS, the $m$-th antenna corresponds to the array element at row $M_{m}^\mathrm{r}$ and column $M_{m}^\mathrm{c}$ with $M_{m}^\mathrm{r} \in \mathcal{M_\mathrm{r}} = \{0,\cdots,M_\mathrm{r}-1\}$, $M_{m}^\mathrm{c} \in \mathcal{M_\mathrm{c}} = \{0,\cdots,M_\mathrm{c}-1\}$, and $m \in \mathcal{M} = \{1,\cdots,M\}$. Then, it follows that $M_{m}^{\mathrm{r}} = \left\lfloor {\left( {{m} - 1} \right)/M_{\mathrm{c}}} \right\rfloor, m \in \mathcal{M}$ and $M_{m}^{\mathrm{c}} = {m} - M_{\mathrm{c}} M_{{m}}^{\mathrm{r}} - 1, m \in \mathcal{M}$. Let ${{\mathbf{p}}_{\mathrm{B},1}} = \left( {x_{\mathrm{B,1}}},{y_{\mathrm{B},1}},0 \right)^T$ denote the coordinates of the first BS antenna. Accordingly, the coordinates of the $m$-th BS antenna are expressed as ${{\mathbf{p}}_{\mathrm{B},m}} = {{\mathbf{p}}_{{\mathrm{B}},1}} + M_{m}^{\mathrm{c}}\tilde l{{\mathbf{m}}_{\mathrm{B}}^{\mathrm{c}}} + M_{m}^{\mathrm{r}}\tilde l{{\mathbf{m}}_{\mathrm{B}}^{\mathrm{r}}}, m \in \mathcal{M}$,
where $\tilde l$ denotes the inter-spacing of the antennas at the BS, $\mathbf{m}_\mathrm{B}^{\mathrm{r}} = \left(0,1,0\right)^T$ and $\mathbf{m}_\mathrm{B}^{\mathrm{c}} = \left(-1,0,0\right)^T$ are the unit vectors. 



Assuming that all the links follow the general Rician fading channels, the BS-IRS 1 channel is modeled as $\mathbf{G} = \sqrt{\kappa_\mathrm{G} / (\kappa_\mathrm{G} + 1)} \mathbf{G}^{\mathrm{LoS}} + \sqrt{1 / (\kappa_\mathrm{G} + 1)} \mathbf{G}^{\mathrm{NLoS}}$, where $\kappa_\mathrm{G}$ denotes the Rician fading factor. Let $\beta$ denote the channel power gain at a reference distance of 1 meter (m). The NLoS component $\bm{G}^{\mathrm{NLoS}}$ has independent entries $\bm{G}^{\mathrm{NLoS}}_{n_1,m} \sim \frac{{\sqrt \beta }}{{{t_{{n_1,m}}}}} \mathcal{CN} \left(0,1\right)$, $\forall n_1 \in \mathcal{N}_1$ and $m \in \mathcal{M}$ with $t_{n_1,m} = \left\| {{{\mathbf{p}}_{\mathrm{B},m}} - {{\mathbf{p}}_{1,n_1}}} \right\|$ denoting the distance between the $m$-th BS antenna and the $n_1$-th element of IRS 1. 
The $(n_1,m)$-th entry of the LoS component is given by
\begin{align}
	\label{G_LoS}
	\left[\bm{G}^{\mathrm{LoS}}\right]_{n_1,m} = \frac{{\sqrt \beta }}{{{t_{{n_1,m}}}}}{e^{ - j\frac{{2\pi }}{\lambda }{t_{{n_1,m}}}}},n_1 \in \mathcal{N}_1,m \in \mathcal{M},
\end{align}
where $\lambda$ denotes the carrier wavelength. Let $d_{n_2,n_1} = \left\| {{\mathbf{p}}_{2,n_2} - {{\mathbf{p}}_{1,n_1}}} \right\|$ and $r_{n_2} = \left\| {{\mathbf{p}}_{\mathrm{U}} - {{\mathbf{p}}_{2,n_2}}} \right\|$ denote the distance from $n_2$-th element of IRS 2 to $n_1$-th element of IRS 1 and to the user, respectively. The channel between IRS 1 and IRS 2, $\mathbf{S} \in \mathbb{C}^{N_2 \times N_1}$, as well as that between IRS 2 and the user, $\mathbf{f} \in \mathbb{C}^{N_2 \times 1}$, can be defined similarly to $\mathbf{G}$, with the corresponding Rician factors $\kappa_\mathrm{S}$ and $\kappa_\mathrm{f}$. As such, the received signal at the user is given by
\begin{align}
	y = {{\mathbf{f}}^H}{{\mathbf{\Phi }}_2}{\mathbf{S}}{{\mathbf{\Phi }}_1}{\mathbf{G}}{\mathbf{w}}s + n_0, 
\end{align}
where $\mathbf{w} \in \mathbb{C}^{M \times 1}$ denotes the transmit beamforming vector of the BS with power constraint $\|\mathbf{w}\|^2 \le P_\mathrm{t}$, and $n_0 \sim \mathcal{CN} \left(0,\sigma_0^2\right)$ represents the additive Gaussian white noise (AWGN) with power $\sigma_0^2$. The transmitted symbol is denoted by $s \in \mathbb{C}$ with unit power. As such, the received SNR is expressed as
\begin{align}
	\label{SNR1}
	\varsigma = \left|{{\mathbf{f}}^H}{{\mathbf{\Phi }}_2}{\mathbf{S}}{{\mathbf{\Phi }}_1}{\mathbf{G}}{\mathbf{w}}\right|^2/\sigma_0^2.
\end{align}

For IRS 1, the basis vectors of its local CCS expressed in the global CCS are given by ${{\mathbf{e}}_{1,\mathrm{x}}} =  (\cos {\theta _1}\sin {\varphi _1},\sin {\theta _1}\sin {\varphi _1}, - \cos {\varphi _1} )^T$, ${{\mathbf{e}}_{1,\mathrm{y}}} = (  - \sin {\theta _1},\cos {\theta _1},0 )^T$, and ${{\mathbf{e}}_{1,\mathrm{z}}} = {\left( {\cos {\theta _1}\cos {\varphi _1},\sin {\theta _1}\cos {\varphi _1},\sin {\varphi _1}} \right)^T}$.
Then, the rotation matrix that maps the local CCS of IRS 1 to the global CCS is given by
\begin{align}
	\label{Q1}
	\!\! {{\mathbf{Q}}_1} ( \mathbf{\Omega}_1 ) \!=& ( {{{\mathbf{e}}_{1,\mathrm{x}}},{{\mathbf{e}}_{1,\mathrm{y}}},{{\mathbf{e}}_{1,\mathrm{z}}}} ) \!=\! \left[ {\begin{array}{*{20}{c}}
			{q_{11}}&{q_{12}}&{q_{13}}\\
			{q_{14}}&{q_{15}}&{q_{16}}\\
			{q_{17}}&0&{q_{18}}
	\end{array}} \right] \nonumber\\
	=&\! \left[  {\begin{array}{*{20}{c}}
			{\cos {\theta _1}\sin {\varphi _1}}&{ - \sin {\theta _1}}&{\cos {\theta _1}\cos {\varphi _1}} \\ 
			{\sin {\theta _1}\sin {\varphi _1}}&{\cos {\theta _1}}&{\sin {\theta _1}\cos {\varphi _1}}\\ 
			{ - \cos {\varphi _1}}&0&{\sin {\varphi _1}}
	\end{array}}  \right] .
\end{align}
Based on \eqref{Q1}, the BS and the IRS 2 coordinates represented in the local CCS of IRS 1 are given by 
\begin{align}
	{\mathbf{p}}_{\mathrm{B}}^{{\mathrm{L1}}} =& {\mathbf{Q}}_1^T\left( \mathbf{\Omega}_1 \right)\left( {{{\mathbf{p}}_{\mathrm{B},1}} - {{\mathbf{p}}_{1,1}}} \right) = {( {x_{\mathrm{B}}^{{\mathrm{L1}}},y_{\mathrm{B}}^{{\mathrm{L1}}},z_{\mathrm{B}}^{{\mathrm{L1}}}} )^T},\\
	{\mathbf{p}}_2^{{\mathrm{L1}}} =& {\mathbf{Q}}_1^T ( \mathbf{\Omega}_1  )\left( {{{\mathbf{p}}_{2,1}} - {{\mathbf{p}}_{1,1}}} \right) = ( {x_2^{{\mathrm{L1}}},y_2^{{\mathrm{L1}}},z_2^{{\mathrm{L1}}}} )^T.
\end{align}
Similarly, the rotation matrix of IRS 2 is given by
\begin{align}
	\label{Q2}
	\!\! {{\mathbf{Q}}_2} ( \mathbf{\Omega}_2 ) \!=& ( {{{\mathbf{e}}_{2,\mathrm{x}}},{{\mathbf{e}}_{2,\mathrm{y}}},{{\mathbf{e}}_{2,\mathrm{z}}}} )
	\!=\! \left[ {\begin{array}{*{20}{c}}
			{q_{21}}&{q_{22}}&{q_{23}}\\
			{q_{24}}&{q_{25}}&{q_{26}}\\
			{q_{27}}&0&{q_{28}}
	\end{array}} \right]\nonumber\\
	=&\! \left[ \!\! {\begin{array}{*{20}{c}}
			{\cos {\theta _2}\sin {\varphi _2}}&{\sin {\theta _2}}&{\cos {\theta _2}\cos {\varphi _2}}\\
			{ - \sin {\theta _2}\sin {\varphi _2}}&{\cos {\theta _2}}&{ - \sin {\theta _2}\cos {\varphi _2}}\\
			{ - \cos {\varphi _2}}&0&{\sin {\varphi _2}}
	\end{array}} \!\! \right],
\end{align}
where ${\mathbf{e}}_{2,\mathrm{x}} = (\cos {\theta _2}\sin {\varphi _2}, - \sin {\theta _2}\sin {\varphi _2}, - \cos {\varphi _2})^T$, ${\mathbf{e}}_{2,\mathrm{y}} = ( \sin {\theta _2},\cos {\theta _2},0 )^T$, ${\mathbf{e}}_{2,\mathrm{z}} = ( \cos {\theta _2}\cos {\varphi _2}, - \sin {\theta _2}\cos {\varphi _2},\sin {\varphi _2} )^T$. Based on \eqref{Q2}, the coordinates of IRS 1 and the user in the local CCS of IRS 2 can be expressed as  ${\mathbf{p}}_1^{{\mathrm{L2}}} \!=\! {\mathbf{Q}}_2^T ( \mathbf{\Omega}_2 )( {{{\mathbf{p}}_{1,1}} - {{\mathbf{p}}_{2,1}}} ) \!=\! ( {x_{\mathrm{1}}^{{\mathrm{L2}}},y_{\mathrm{1}}^{{\mathrm{L2}}},z_{\mathrm{1}}^{{\mathrm{L2}}}} )^T$ and ${\mathbf{p}}_{\mathrm{U}}^{{\mathrm{L2}}} = {\mathbf{Q}}_2^T( \mathbf{\Omega}_2 )( {{{\mathbf{p}}_{\mathrm{U}}} - {{\mathbf{p}}_{2,1}}} ) = ( {x_{\mathrm{U}}^{{\mathrm{L2}}},y_{\mathrm{U}}^{{\mathrm{L2}}},z_{\mathrm{U}}^{{\mathrm{L2}}}} )^T$, respectively.

We consider the far-field scenario, assume that all IRS elements share the same reflection coefficient, and approximate $t_{n_1,m} \approx t_{1,1}$, $d_{n_2,n_1} \approx d_{1,1}$, and $r_{n_2} \approx r_{1}$ for the calculation of channel amplitudes. Accordingly, the effective aperture gain at the IRS 1 and IRS 2 are given by \cite{rotate24}
\begin{align}
	\bar F\left( \mathbf{\Omega}_1 \right) = \cos {{\bar \phi }^{\mathrm{i}}} \cos {{\bar \phi }^{\mathrm{r}}}, \;\;
	\tilde F\left( \mathbf{\Omega}_2 \right) = \cos {{\tilde \phi }^{\mathrm{i}}} \cos {{\tilde \phi }^{\mathrm{r}}},
\end{align}
respectively, where the incident and reflected elevation angles of IRS 1 are given by
\small
\begin{align}
	{{\bar \phi }^{\mathrm{i}}} \!\!\!=&\! \arccos \frac{{z_{\mathrm{B}}^{{\mathrm{L1}}}}}{{\left\| {{{\mathbf{p}}_{\mathrm{B},1}} - {{\mathbf{p}}_{1,1}}} \right\|}} \nonumber\\
	\!\!\!=&\! \arccos \! \frac{{{( {{x_{\mathrm{B},1}} - {x_{1,1}}} )}q_{13} + ( {{y_{\mathrm{B},1}} - {y_{1,1}}} )q_{16} - {  {z_{1,1}}} q_{18}}}{{\sqrt {(x_{\mathrm{B},1} - x_{1,1})^2 + {{( {{y_{\mathrm{B},1}} - {y_{1,1}}} )}^2} + z_{1,1}^2} }}, \\
	{{\bar \phi }^{\mathrm{r}}} \!\!\!=&\! \arccos \frac{{z_2^{{\mathrm{L1}}}}}{{\left\| {{{\mathbf{p}}_{2,1}} - {{\mathbf{p}}_{1,1}}} \right\|}} \nonumber\\
	\!\!\!=&\! \arccos \! \frac{{( {{x_{2,1}} \!-\! {x_{1,1}}} )q_{13}  \!\!+\!\! ( {{y_{2,1}} \!-\! {y_{1,1}}} )q_{16} \!\!+\!\! ( {{z_{2,1}} \!-\! {z_{1,1}}} ) q_{18}}}{{\sqrt {( {{x_{1,1}} \!-\! {x_{2,1}}} )^2 \!+\! {( {{y_{2,1}} \!-\! {y_{1,1}}} )^2} \!+\! {( {{z_{2,1}} \!-\! {z_{1,1}}} )^2}} }}. 
\end{align}
\normalsize
Similarly, the incident and reflected elevation angles of IRS 2 are rewritten as
\small
\begin{align}
	{{\tilde \phi }^{\mathrm{i}}} \!\!\!=& \arccos \frac{{z_{\mathrm{1}}^{{\mathrm{L2}}}}}{{\left\| {{{\mathbf{p}}_{1,1}} - {{\mathbf{p}}_{2,1}}} \right\|}} \nonumber\\
	\!\!\!=&\! \arccos \! \frac{{( {{x_{1,1}} \!-\! {x_{2,1}}} )q_{23} \!\!+\!\! ( {{y_{1,1}} \!-\! {y_{2,1}}} )q_{26} \!\!+\!\! ( {{z_{1,1}} \!-\! {z_{2,1}}} )q_{28}}}{{\sqrt {( {{x_{1,1}} \!-\! {x_{2,1}}} )^2 \!+\! {{( {{y_{1,1}} \!-\! {y_{2,1}}} )}^2} \!+\! {{( {{z_{1,1}} \!-\! {z_{2,1}}} )}^2}} }}, \label{phi_i_tilde}\\
	{{\tilde \phi }^{\mathrm{r}}} \!\!\!=& \arccos \frac{{z_{\mathrm{U}}^{{\mathrm{L2}}}}}{{\left\| {{{\mathbf{p}}_{\mathrm{U}}} - {{\mathbf{p}}_{2,1}}} \right\|}}\nonumber\\
	\!\!\!=&\! \arccos \! \frac{{( {{x_{\mathrm{U}}} - {x_{2,1}}} )q_{23} \!+\! ( {{y_{\mathrm{U}}} - {y_{2,1}}} )q_{26} \!-\! {z_{2,1}}q_{28}}}{{\sqrt {{{( {{x_{\mathrm{U}}} - {x_{2,1}}} )}^2} + {{( {{y_{\mathrm{U}}} - {y_{2,1}}} )}^2} + z_{2,1}^2} }}.\label{phi_r_tilde}	
\end{align}
\normalsize

As such, the received signal at the user can be rewritten as $y = \sqrt{\bar F\left(\bm{\Omega}_1\right)\tilde F\left(\bm{\Omega}_2\right)}{{\mathbf{f}}^H}{{\tilde{\mathbf{ \Phi }}}}{\mathbf{S}}{{\bar{\mathbf{ \Phi }}}}{\mathbf{G}}{\mathbf{w}}s + n_0$,
where ${{\bar{\mathbf{ \Phi }}}} = \operatorname{diag}({e^{j{{\bar \psi }_{1}}}},\cdots, {e^{j{{\bar \psi }_{{N_1}}}}})$ and ${{\tilde{\mathbf{ \Phi }}}} = \operatorname{diag}({e^{j{{\tilde \psi }_{1}}}},\cdots, {e^{j{{\tilde \psi }_{{N_2}}}}})$.
The received SNR is given by
\begin{align}
	\label{SNR2}
	\varsigma = \bar F\left( \mathbf{\Omega}_1 \right)\tilde F\left( \mathbf{\Omega}_2 \right)  |{{\mathbf{f}}^H}{{\tilde{\mathbf{ \Phi }}}}{\mathbf{S}}{{\bar{\mathbf{ \Phi }}}}{\mathbf{G}}{\mathbf{w}} |^2 / \sigma _0^2.
\end{align}

We denote the unit direction vectors as $\bar{\mathbf{a}}_{\mathrm{t}} = \frac{{{{\mathbf{p}}_{1,1}} - {{\mathbf{p}}_{\mathrm{B},1}}}}{{\left\| {{{\mathbf{p}}_{1,1}} - {{\mathbf{p}}_{\mathrm{B},1}}} \right\|}}$, ${{{\bar{\mathbf{a}}}}_{\mathrm{r}}} = {{{\tilde{\mathbf{a}}}}_{\mathrm{t}}} = \frac{{{{\mathbf{p}}_{2,1}} - {{\mathbf{p}}_{1,1}}}}{{\left\| {{{\mathbf{p}}_{2,1}} - {{\mathbf{p}}_{1,1}}} \right\|}}$, and ${{{\tilde{\mathbf{a}}}}_{\mathrm{r}}} = \frac{{{{\mathbf{p}}_{\mathrm{U}}} - {{\mathbf{p}}_{2,1}}}}{{\left\| {{{\mathbf{p}}_{\mathrm{U}}} - {{\mathbf{p}}_{2,1}}} \right\|}}$, where $\bar{\mathbf{a}}_{\mathrm{t}}$ and ${{{\bar{\mathbf{a}}}}_{\mathrm{r}}}$ stand for the incident and reflected
directions at IRS 1, while ${{{\tilde{\mathbf{a}}}}_{\mathrm{t}}}$ and ${{{\tilde{\mathbf{a}}}}_{\mathrm{r}}}$ denote the incident and reflected
directions at IRS 2, respectively. For effective reflection, the BS and IRS 2 must lie on the reflective side of the IRS 1, while the IRS 1 and user must lie on the reflective side of the IRS 2, which leads to ${\bar{\mathbf{a}}_{\mathrm{t}}^T{{\mathbf{k}_1}}} \le 0$, ${\bar{\mathbf{a}}_{\mathrm{r}}^T{{\mathbf{k}_1}}} \ge 0$, ${\tilde{\mathbf{a}}_{\mathrm{t}}^T{{\mathbf{k}_2}}} \le 0$, ${\tilde{\mathbf{a}}_{\mathrm{r}}^T{{\mathbf{k}_2}}} \ge 0$. Then, it follows that $z_\mathrm{B}^\mathrm{L1} \ge 0$, $z_2^\mathrm{L1} \ge 0$, $z_1^\mathrm{L2} \ge 0$, and $z_\mathrm{U}^\mathrm{L2} \ge 0$, i.e.,
\small
\begin{align}
	& {{{x_{\mathrm{B},1}}\cos {\theta _1}\cos {\varphi _1} \!\!+\!\! ( {{y_{\mathrm{B},1}} \!\!-\!\! {y_{1,1}}} ) \! \sin {\theta _1}\cos {\varphi _1} \!\!-\!\! {  {z_{1,1}}} \sin {\varphi _1}}} \ge 0 , \label{Bat1}\\
	& {{( {{y_{2,1}} - {y_{1,1}}} )\sin {\theta _1}\cos {\varphi _1} + ( {{z_{2,1}} - {z_{1,1}}} ) \sin {\varphi _1}}} \ge 0 , \label{2at1}\\
	& {{ - ( {{y_{1,1}} - {y_{2,1}}} )\sin {\theta _2}\cos {\varphi _2} + ( {{z_{1,1}} - {z_{2,1}}} )\sin {\varphi _2}}} \ge 0 , \label{1at2}\\
	& ( {{x_{\mathrm{U}}} - {x_{2,1}}} ) \cos {\theta _2}  \cos {\varphi _2} - ( {{y_{\mathrm{U}}} - {y_{2,1}}} )\sin {\theta _2}\cos {\varphi _2} \nonumber\\
	&- {z_{2,1}}\sin {\varphi _2} \ge 0. \label{Uat2}
\end{align}
\normalsize

We aim to maximize the received SNR by jointly optimizing the phase shifts and the rotation of the two IRSs, as well as the BS beamforming vector. The optimization problem can be formulated as 
\begin{subequations}
	\label{pro:area}
	\begin{align}
		\mathop {\max }\limits_{\bar{\mathbf{ \Phi }},\tilde{\mathbf{ \Phi }},\mathbf{w},\mathbf{\Omega}_1, \mathbf{\Omega}_2} \;\;\;& \varsigma \\
		\mathrm{s.t.} \;\;\;\;\;\;\;\;\; 
		& \|\mathbf{w}\|^2 \le P_\mathrm{t}, \\
		& |[\bar{\mathbf{ \Phi }}]_{n_1,n_1}| = 1, \forall n_1 \in \mathcal{N}_1, \\
		& |[\tilde{\mathbf{ \Phi }}]_{n_2,n_2}| = 1, \forall n_2 \in \mathcal{N}_2, \\
		& \eqref{Bat1}, \eqref{2at1}, \eqref{1at2}, \eqref{Uat2}.
	\end{align}
\end{subequations}
The main challenges for solving problem \eqref{pro:area} are non-convex objective function, non-convex constraints, and highly-coupled optimization variables. To overcome these issues, we consider this problem under the LoS and general Rician channel cases by decomposing the original problem into sub-problems, which enables tractable optimization w.r.t. the involved variables. The details are presented in the following section.

\section{Proposed Solutions}
In this section, we first investigate the SNR maximization problem in the LoS channel case. To draw useful insights, we obtain the optimal azimuth angle for a particular case with $z_{1,1} = z_{2,1} = 0$. Then, two frameworks are proposed under the general Rician fading channels. 

\subsection{LoS channel}
\label{LOScase}
We first consider the LoS channel case with $\kappa_\mathrm{G} \to \infty$, $\kappa_\mathrm{S} \to \infty$ and $\kappa_\mathrm{f} \to \infty$. According to \cite{double_IRS}, the channel matrix $\mathbf{G^\mathrm{LoS}}$ and $\mathbf{S^\mathrm{LoS}}$ can be decomposed as the product of two signature vectors $\mathbf{g}_1$ and $\mathbf{g}_2$ as well as $\mathbf{s}_1$ and $\mathbf{s}_2$, respectively. The optimized phase shifts of IRS 1 and IRS 2, as well as the optimized transmit beamforming vector are given by $\bar \psi _{n_1}^\star = -\arg([\mathbf{s}_1]_{n_1}) -\arg([\mathbf{g}_2]_{n_1})$, $n_1 \in \mathcal{N}_1$, $\tilde \psi _{n_2}^\star = \arg([\mathbf{f}^{\mathrm{LoS}}]_{n_2}) - \arg([\mathbf{s}_2]_{n_2})$, $n_2 \in \mathcal{N}_2$, and $\mathbf{w}^\star = \sqrt{P_\mathrm{t}}\frac{(\mathbf{g}_1^*)}{||\mathbf{g}_1||}$, respectively.
Substituting $\mathbf{w}^{\star}$, $\bar{\mathbf{ \Phi }}^{\star}$, and $\tilde{\mathbf{ \Phi }}^{\star}$ into \eqref{SNR2} yields 
\begin{align}
	\label{SNR3}
	\varsigma = \bar F\left( \mathbf{\Omega}_1 \right)\tilde F\left( \mathbf{\Omega}_2 \right) \frac{P_\mathrm{t} \beta^3 N_1^2 N_2^2 M}{\sigma _0^2 t_{1,1}^2 d_{1,1}^2 r_1^2}.
\end{align}
From \eqref{SNR3}, it can be readily verified that optimally equipping IRS 1 and IRS 2 with the same number of reflecting elements, i.e., $N_1 = N_2 = N/2$, yields a quartic power scaling w.r.t. $N$ \cite{double_IRS}, given by $\varsigma = \bar F\left( \mathbf{\Omega}_1 \right)\tilde F\left( \mathbf{\Omega}_2 \right) \frac{P_\mathrm{t} \beta^3 N^4 M}{16\sigma _0^2 t_{1,1}^2 d_{1,1}^2 r_1^2}$. Despite the severe path loss of the cascaded link, the proposed double-IRS architecture remains feasible for realistic blocked regions. The quartic cooperative passive beamforming gain plays a critical role in compensating for the triple-hop attenuation.

For any given $\mathbf{\Omega}_2$, problem \eqref{pro:area} is equivalent to
\begin{align}
	\label{pro:rotate_1}
	\mathop {\max }\limits_{\mathbf{\Omega}_1} \; \bar F\left( \mathbf{\Omega}_1 \right) \;\;\;\; \mathrm{s.t.} \; \eqref{Bat1}, \eqref{2at1}.
\end{align}

\begin{Proposition}
	\label{pro:rotation}
	The optimal azimuth and elevation angles of IRS 1 are given by
	\begin{align}
		\label{optimal1}
		\theta_1^* = \arctan \frac{\Delta_{1,\mathrm{y}}}{\Delta_{1,\mathrm{x}}}, \varphi _1^*=\arcsin \frac{\Delta_{1,\mathrm{z}}}{||\mathbf{\Delta}_1||},
	\end{align}
	where $\mathbf{\Delta}_1 = (\Delta_{1,\mathrm{x}},\Delta_{1,\mathrm{y}},\Delta_{1,\mathrm{z}})^T = -\bar{\mathbf{a}}_{\mathrm{t}}+\bar{\mathbf{a}}_{\mathrm{r}}$.
\end{Proposition}
{\it{Proof:}} Please refer to Appendix A. ~$\hfill\blacksquare$

For any given $\mathbf{\Omega}_1$, problem \eqref{pro:area} is equivalent to
\begin{align}
	\label{optimal2}
	\mathop {\max }\limits_{\mathbf{\Omega}_2} \; \bar F\left( \mathbf{\Omega}_2 \right) \;\;\;\; \mathrm{s.t.} \; \eqref{1at2}, \eqref{Uat2}.
\end{align}
Similar to Proposition \ref{pro:rotation}, the optimal azimuth and elevation angles of IRS 2 can be expressed as 
\begin{align}
	\theta_2^* = \arctan \frac{\Delta_{2,\mathrm{y}}}{\Delta_{2,\mathrm{x}}}, \varphi _2^*=\arcsin \frac{\Delta_{2,\mathrm{z}}}{||\mathbf{\Delta}_2||},
\end{align}
where $\mathbf{\Delta}_2 = (\Delta_{2,\mathrm{x}},\Delta_{2,\mathrm{y}},\Delta_{2,\mathrm{z}})^T = -\tilde{\mathbf{a}}_{\mathrm{t}}+\tilde{\mathbf{a}}_{\mathrm{r}}$.

The results reveal that the rotation designs of the two IRSs depend on their propagation geometry. Specifically, the optimal normal vector of each IRS is aligned with the direction determined by the difference between the incident and reflected propagation directions, which is consistent with the specular reflection law. This implies that the optimal IRS orientation is determined by the geometry of the incident and reflected links, thereby maximizing the effective aperture gain. Moreover, under the special case with $z_{1,1}= z_{1,2} = 0$, we have $\Delta_{1,z} = \Delta_{2,z} = 0$, the optimal elevation angles are given by $\varphi _1^* = \varphi _2^* = 0$. It shows that the azimuth angle optimization suffices to achieve the optimal performance. This result provides useful insights for practical hardware implementation, indicating that single-axis rotation may be sufficient in certain scenarios, thereby significantly simplifying the mechanical design of rotatable IRSs. 

\subsection{Rician Fading Channel}
To address the general Rician fading channel case, we develop two efficient frameworks for solving problem \eqref{pro:area}, depending on the timescale of the IRS rotation design.

\textbf{Framework 1:} Motivated by the practical latency of mechanical rotation, we consider a rotation design based on long-term statistical CSI. Specifically, the orientations of the two IRSs are optimized using the long-term channel geometry determined by the LoS components, which can be derived from the previous subsection. Given the optimized IRS rotations, the optimized transmit beamforming at the BS $\mathbf{w}$ and the passive beamforming at the two IRSs, $\bar{\mathbf{ \Phi }}$ and $\tilde{\mathbf{ \Phi }}$, are jointly optimized in close form based on instantaneous CSI. This method is computationally efficient and practically appealing.

\textbf{Framework 2:} To characterize the potential performance gain of joint rotation and beamforming optimization, we further consider a more general scheme where the IRS rotations are optimized using a PSO-based method. In this case, we optimize the IRS rotations $\mathbf{\Omega}_1$ and $\mathbf{\Omega}_2$ via the PSO-based method. By updating $\mathbf{\Omega}_1$, $\mathbf{\Omega}_2$, $\mathbf{w}$, $\bar{\mathbf{ \Phi }}$, and $\tilde{\mathbf{ \Phi }}$ in an iterative manner until convergence is achieved, we can obtain a high-quality solution to the original optimization problem. This joint optimization scheme can better adapt to channel conditions and thus potentially achieve higher performance. The corresponding PSO-based rotation design are provided in the following. 

\subsubsection{IRS Rotation Optimization}
The rotation optimization problems for IRS 1 and IRS 2 have a similar formulation. Therefore, we focus on the rotation design of IRS 1 in the following, whereas that of IRS 2 is omitted for brevity. 

Due to the inherent non-convexity of $\bar F\left( \mathbf{\Omega}_1 \right)$ and the nonlinear dependence of $z_{\mathrm{B}}^{\mathrm{L1}}$ and $z_2^{\mathrm{L1}}$ on $\mathbf{\Omega}_1$, analytical optimization or gradient-based methods are generally ineffective. Moreover, the feasible domain of $\mathbf{\Omega}_1 = (\theta_1, \varphi_1)^T$ is compact but highly irregular. These motivate us that, for any given $\bar{\mathbf{ \Phi }}$, $\tilde{\mathbf{ \Phi }}$, and $\mathbf{w}$, the rotation angles of the two IRSs can be optimized via the PSO-based method. Each particle in the PSO swarm represents a candidate rotation angle $\mathbf{\Omega}_1$, and its performance is measured by the penalized fitness function. For example, the penalized fitness function w.r.t. $\mathbf{\Omega}_1$ is given by
\begin{align}
	\label{fitness}
	\mathcal{L} (\mathbf{\Omega}_1 ) = \varsigma - \tau (\max  \{0, -z_{\mathrm{B}}^{\mathrm{L1}} \} + \max  \{0, -z_{\mathrm{2}}^{\mathrm{L1}} \} ).
\end{align}
where $\tau>0$ guarantees that constraints \eqref{Bat1} and \eqref{2at1} are satisfied by penalizing infeasible particle positions. The position and velocity of the $b$-th particle are updated at the $(t+1)$-th iteration as $\mathbf{\Omega}_b^{ (t+1 )} = \mathcal{G} (\mathbf{\Omega}_b^{ (t )} + {\bm{\nu}}_b^{ (t )} )$, $\bm{\nu}_b^{ (t+1 )} = \omega^{ (t )}\bm{\nu}_b^{ (t )} + c_1  \upsilon _1  (\mathbf{\Omega}_{b,\mathrm{lbest}} - \mathbf{\Omega}_b^{ (t )})+ c_2  \upsilon _2  (\mathbf{\Omega}_{\mathrm{gbest}} - \mathbf{\Omega}_b^{ (t )})$,
where $c_1$ and $c_2$ denote the cognitive and social coefficients, $\upsilon _1$ and $\upsilon _2$ are random variables with a uniform distribution over $\left[0,1\right]$, $\mathbf{\Omega}_{b,\mathrm{lbest}}$ and $\mathbf{\Omega}_{\mathrm{gbest}}$ denote the personal and global best solutions, respectively. The projection operator $\mathcal{G}\left(\cdot\right)$ maps each entry of $\mathbf{\Omega}_b$ to the feasible interval. The inertia weight is denoted by $\omega^{\left(t\right)} = \left(\omega_\mathrm{i} - \omega_\mathrm{e}\right) \left({T_\mathrm{max}-t}\right)/{T_\mathrm{max}} + \omega_\mathrm{e}$ where $T_\mathrm{max}$, $\omega_\mathrm{i}$, and $\omega_\mathrm{e}$ stand for total number of iterations, the initial and the final inertia weight. The computational complexity of the PSO-based algorithm is $\mathcal{O}\left(2BT_{\mathrm{max}}\right)$, where $B$ is the total number of particles.

For both frameworks, the beamforming design are optimized as follows.
\subsubsection{BS Beamforming Optimization}
For any given $\bar{\mathbf{ \Phi }}$ and $\tilde{\mathbf{ \Phi }}$, the BS beamforming optimization problem is formulated as
\begin{align}
	\label{pro:BS}
	\mathop {\max }\limits_{\mathbf{w}} \;  |{{\mathbf{f}}^H}{{\tilde{\mathbf{ \Phi }}}}{\mathbf{S}}{{\bar{\mathbf{ \Phi }}}}{\mathbf{G}}{\mathbf{w}} |^2 
	\;\;\; \mathrm{s.t.} \;
	\|\mathbf{w}\|^2 \le P_\mathrm{t}.
\end{align}
It is readily verified that the maximum ratio transmission (MRT) is the optimal transmit beamforming solution to problem \eqref{pro:BS}, i.e., $\mathbf{w}^{\star} = \sqrt{P_\mathrm{t}} \frac{({{\mathbf{f}}^H}{{\tilde{\mathbf{ \Phi }}}}{\mathbf{S}}{{\bar{\mathbf{ \Phi }}}}{\mathbf{G}})^H}{\|{{\mathbf{f}}^H}{{\tilde{\mathbf{ \Phi }}}}{\mathbf{S}}{{\bar{\mathbf{ \Phi }}}}{\mathbf{G}}\|}$.

\subsubsection{IRS 1 Beamforming Optimization}
For any given $\mathbf{w}$ and $\tilde{\mathbf{ \Phi }}$, we optimize the beamforming vector of IRS 1, for which problem \eqref{pro:area} is reduced to
\begin{align}
	\label{pro:IRS1}
	\mathop {\max }\limits_{\bar{\mathbf{ \Phi }}} \; \left|{{\mathbf{f}}^H}{{\tilde{\mathbf{ \Phi }}}}{\mathbf{S}}{{\bar{\mathbf{ \Phi }}}}{\mathbf{G}}{\mathbf{w}}\right|^2 \;\;\;
	\mathrm{s.t.} \;
	|[\bar{\mathbf{ \Phi }}]_{n_1,n_1}| = 1, \forall n_1 \in \mathcal{N}_1.
\end{align}
Let $\tilde{\mathbf{h}}^H = {{\mathbf{f}}^H}{{\tilde{\mathbf{ \Phi }}}}{\mathbf{S}} \in \mathbb{C}^{1 \times N_1}$ and $\mathbf{g} = {\mathbf{G}}{\mathbf{w}}  \in \mathbb{C}^{N_1 \times 1}$. The optimal phase shifts of the IRS 1 are given by $\bar \psi _{n_1}^\star = \arg (\tilde{\mathbf{h}}_{n_1}) - \arg (\mathbf{g}_{n_1}), n_1 \in \mathcal{N}_1$.

\subsubsection{IRS 2 Beamforming Optimization}
For any given $\mathbf{w}$ and $\bar{\mathbf{ \Phi }}$, we optimize the beamforming vector of IRS 2, for which problem \eqref{pro:area} is reduced to
\begin{align}
	\label{pro:IRS2}
	\mathop {\max }\limits_{\tilde{\mathbf{ \Phi }}} \;  |{{\mathbf{f}}^H}{{\tilde{\mathbf{ \Phi }}}}{\mathbf{S}}{{\bar{\mathbf{ \Phi }}}}{\mathbf{G}}{\mathbf{w}} |^2 
	\;\;\; \mathrm{s.t.} \;
	|[\tilde{\mathbf{ \Phi }}]_{n_2,n_2}| = 1, \forall n_2 \in \mathcal{N}_2.
\end{align}
Let $\bar{\mathbf{h}} = {\mathbf{S}}{{\bar{\mathbf{ \Phi }}}}{\mathbf{G}}{\mathbf{w}} \in \mathbb{C}^{N_2 \times 1}$. The optimal phase shifts of the IRS 2 are given by $\tilde \psi _{n_2}^\star = \arg (\mathbf{f}_{n_2}) - \arg (\bar{\mathbf{h}}_{n_2}),  n_2 \in \mathcal{N}_2$.

The AO algorithm in Framework 1 is guaranteed to converge since the BS/IRS beamforming sub-problems yield closed-form globally optimal solutions. The main computational complexity of this algorithm is $\mathcal{O}(I_\mathrm{AO}(N_1 M + N_1 N_2))$, where $I_\mathrm{AO}$ denotes the number of iterations required for convergence. In Framework 2, the AO-PSO algorithm is guaranteed to converge since the PSO method for IRS rotation optimization produces a monotonically non-decreasing and bounded global-best fitness value. Moreover, the closed-form rotation angles obtained in \eqref{optimal1} and \eqref{optimal2} are used as the initial solution for the PSO-based rotation optimization. The corresponding computational complexity is $\mathcal{O}(I_\mathrm{AO}(4B T_\mathrm{max} + N_1 M + N_1 N_2))$.

\section{Numerical Results}
In this section, we present numerical results under the following setup. The locations of the first element of the BS, IRS 1, and IRS 2, as well as the user are $(-7, -15, 0)$ m, $(0, 25, 5)$ m, $(5, -20, 10)$ m, and $(15, 20, 0)$ m. The carrier frequency is 2.4 GHz. Unless otherwise stated, we set $P_\mathrm{t}=30$ dBm, $\sigma_0^2 = -80$ dBm, $\beta = -40$ dB, $M = 64$, $N_1 = N_2 = 256$, $N = N_1 + N_2 = 512$, $l = \tilde l = \lambda /2$, $T_\mathrm{max} = 30$, and $B = 400$. For comparison, we consider the following schemes: 1) single R-IRS: a single rotatable IRS with $N = 512$, placed such that its BS–IRS–user cascaded distance product is nearly the same as that of the double-IRS system; 2) double F-IRS: double IRSs are fixed with $\mathbf{\Omega}_1 = \mathbf{\Omega}_2 = (-\pi/4,-\pi/4)^T$; 3) a single IRS is fixed with rotation angles $\mathbf{\Omega} = (-\pi/4,-\pi/4)^T$.


\begin{figure}[t]
	\centering
	\includegraphics[width=0.35\textwidth]{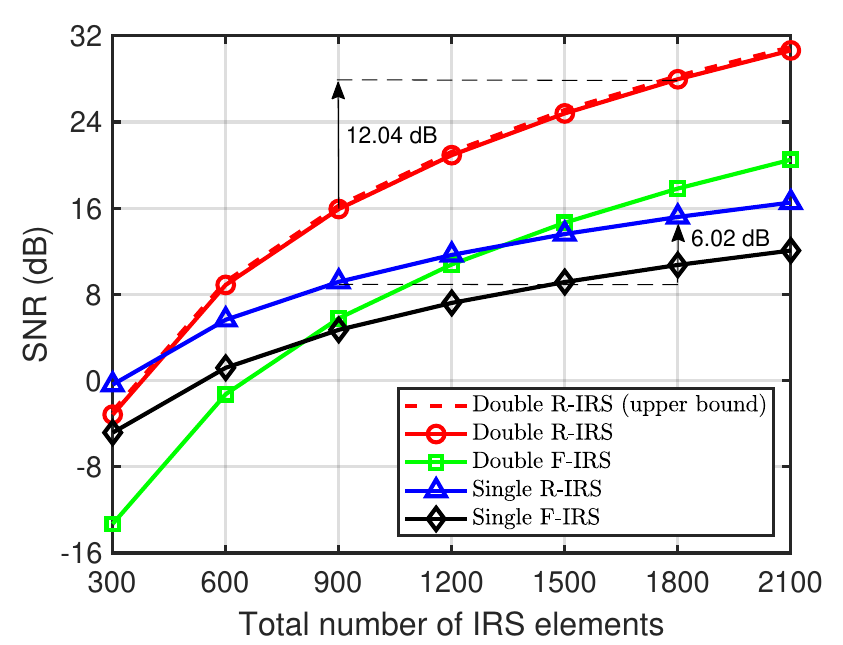}
	\caption{SNR versus total number of IRS elements under the LoS channels.}
	\label{fig:MN}
\end{figure}
Under the LoS channels, we plot the SNR versus the total number of IRS elements in Fig. \ref{fig:MN}. The upper bound corresponds to the ideal case, i.e., $\bar F\left( \mathbf{\Omega}_1 \right) = \tilde F\left( \mathbf{\Omega}_2 \right) = 1$. With the optimized rotation under the R-IRS scheme for $N = 300$ , we have $\bar F\left( \mathbf{\Omega}_1 \right) = 0.9668$ and $\tilde F\left( \mathbf{\Omega}_2 \right) = 0.9657$, resulting in $\bar F\left( \mathbf{\Omega}_1 \right) \tilde F\left( \mathbf{\Omega}_2 \right) = 0.9336$. It can be observed that the proposed design closely approaches the upper bound, which indicates the effectiveness of the proposed rotation design. The received SNR of all the schemes increases with $N$ thanks to the higher passive beamforming gain. Doubling the total number of elements from $N = 900$ to $N = 1800$ yields a 12.04 dB gain for the cooperative double rotatable IRS-aided system. This is expected due to the quartic power scaling order $\mathcal{O}(N^4)$ achieved by cooperative double IRS beamforming, which is consistent with the discussion in Section \ref{LOScase}. When $N$ is sufficiently large, both the double R-IRS and the double F-IRS can achieve better performance than the single R-IRS. Moreover, the required number of IRS elements is reduced by more than 60\% when deploying the double R-IRS instead of the double F-IRS to outperform the single R-IRS, thanks to the additional reflection gain brought by the IRS rotation.

\begin{figure}[t]
	\centering
	\includegraphics[width=0.35\textwidth]{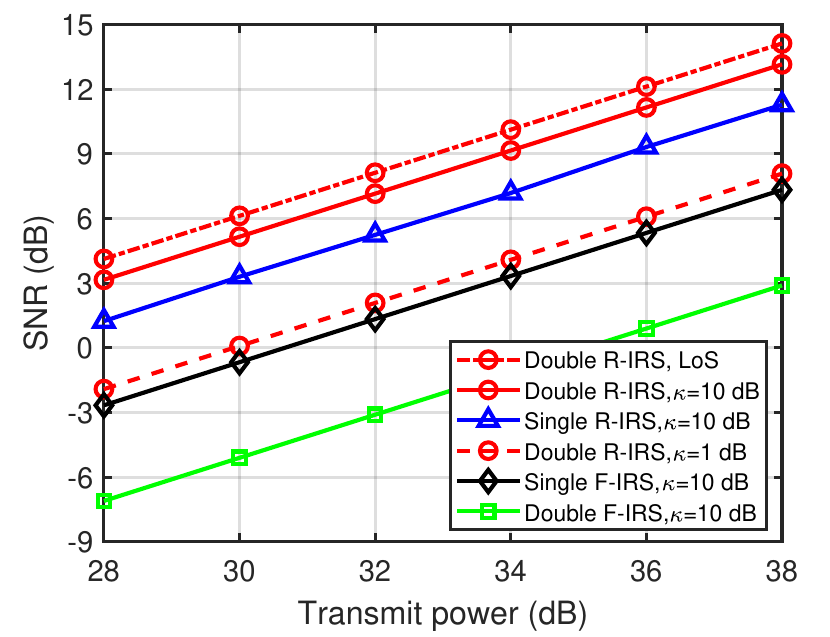}
	\caption{Impact of Rician factor on the performance.}
	\label{fig:Rician}
\end{figure}
In Fig. \ref{fig:Rician}, we plot the SNR versus the transmit power under different Rician fading factors $\kappa = \kappa_\mathrm{G} = \kappa_\mathrm{S} = \kappa_\mathrm{f}$. Three IRS rotation optimization schemes are compared: 1) LoS: the LoS component-based design as presented in Framework 1; 2) PSO: the PSO-based design as presented in Framework 2; 3) GA \cite{GA}: the genetic algorithm-based design. It can be observed that varying the Rician factor has little impact on the performance gap among different rotation optimization schemes, especially when the LoS component becomes dominant, i.e., under large Rician factors. Although the GA-based scheme achieves performance close to that of the PSO-based scheme, the latter is more suitable for the considered rotation optimization problem. Specifically, the IRS rotation variables are continuous and low-dimensional with bounded feasible regions, which allows PSO to efficiently explore the feasible search space through velocity–position updates. Moreover, the closed-form rotation angles derived in \eqref{optimal1} and \eqref{optimal2} provide a high-quality initialization that further facilitates the PSO-based scheme. In contrast, the global search in GA relies on genetic operations such as crossover and mutation, which leads to lower search efficiency. When the Rician factor is -4 dB, the PSO-based rotation design in Framework 2 achieves 0.6 dB higher SNR than the LoS-based rotation design in Framework 1. This is expected since the NLoS components introduce orientation mismatch, which leads to performance degradation. Nevertheless, Framework 1 enjoys the benefits of lower computational complexity and overhead, which may be promising in practical scenarios.

\begin{figure}[t]
	\centering
	\includegraphics[width=0.35\textwidth]{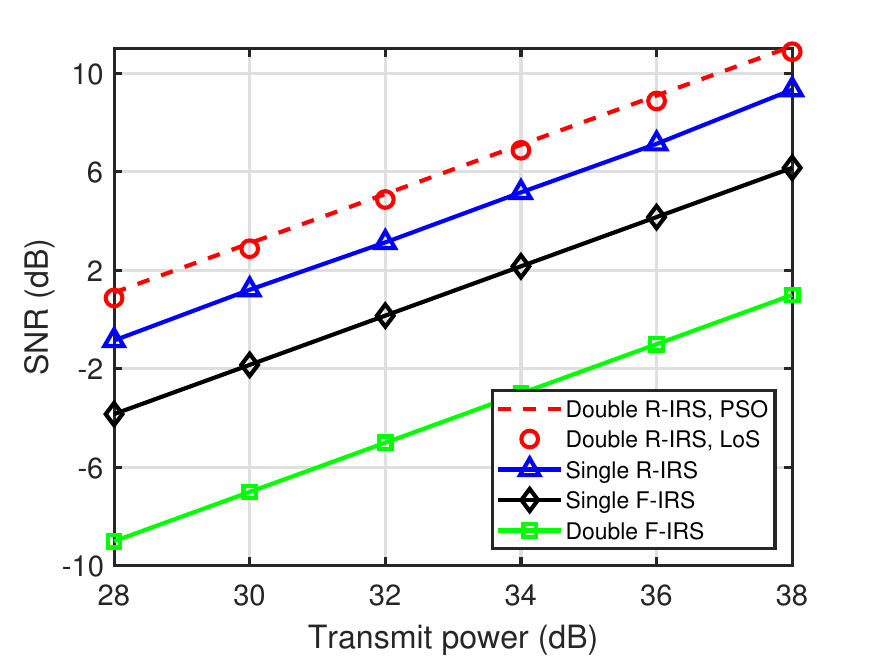}
	\caption{SNR versus the transmit power under the Rician fading channels.}
	\label{fig:1}
\end{figure}

In Fig. \ref{fig:1}, the SNR versus the transmit power is plotted when $\kappa = \kappa_\mathrm{G} = \kappa_\mathrm{S} = \kappa_\mathrm{f} = 5$ dB. It can be observed that the schemes with optimized IRS rotation significantly outperform those with fixed orientation. This is because appropriately adjusting the IRS orientations to align the incident and reflected propagation directions can enhance the effective aperture gain. The performance gain brought by rotation optimization is more pronounced in the double IRS-aided system, since the signal travels through double-reflection links. Orientation mismatch at one of the IRSs may result in a significant loss of the cascaded channel gain. Furthermore, compared with the single R-IRS scheme, the double R-IRS achieves a higher SNR due to the additional cooperative passive beamforming gain provided by the two IRSs. These results highlight the importance of carefully designing the rotation of both IRSs, which enables the system to fully exploit the cooperative beamforming capability of the double IRS architecture.

\section{Conclusion}
This letter studied joint rotation and beamforming optimization for a cooperative double rotatable IRS-aided communication system under a practical angle-dependent reflection model. For the LoS case, the optimal IRS orientations are derived in closed form based on the propagation geometry, where the corresponding optimal IRS normal follows the specular reflection law. For more general Rician fading channels, two efficient rotation frameworks were proposed, namely, a low-complexity geometry-based design and a PSO-based optimization scheme. Numerical results showed that the performance of the former framework is close to that of the latter. Moreover, the results highlighted the importance of carefully designing the orientations of both IRSs and demonstrated the benefits of the cooperative double rotatable IRS in improving SNR under various setups.
Several interesting directions can be further explored in future work, including accounting for the IRS radiation pattern as in \cite{RA2}, investigating near-field effects for large IRSs, and extending the framework to the general case with both single- and double-reflection links.


\section*{Appendix A}
The objective function of problem \eqref{pro:rotate_1} can be rewritten as 
\begin{align}
	\bar F\left( \mathbf{\Omega}_1 \right) &= (-\mathbf{e}_{1,\mathrm{z}}^T \bar{\mathbf{a}}_{\mathrm{t}})(\mathbf{e}_{1,\mathrm{z}}^T \bar{\mathbf{a}}_{\mathrm{r}}) \nonumber\\
	&= ((\mathbf{e}_{1,\mathrm{z}}^T(-\bar{\mathbf{a}}_{\mathrm{t}}+\bar{\mathbf{a}}_{\mathrm{r}}))^2-(\mathbf{e}_{1,\mathrm{z}}^T(\bar{\mathbf{a}}_{\mathrm{t}}+\bar{\mathbf{a}}_{\mathrm{r}}))^2)/4 \nonumber\\
	&\mathop \le \limits^{(a_1)} (\mathbf{e}_{1,\mathrm{z}}^T(-\bar{\mathbf{a}}_{\mathrm{t}}+\bar{\mathbf{a}}_{\mathrm{r}}))^2/4 \nonumber\\
	&\mathop \le \limits^{(a_2)}
	(||\mathbf{e}_{1,\mathrm{z}}||^2||(-\bar{\mathbf{a}}_{\mathrm{t}}+\bar{\mathbf{a}}_{\mathrm{r}})||^2)/4 \nonumber\\
	& = ||(-\bar{\mathbf{a}}_{\mathrm{t}}+\bar{\mathbf{a}}_{\mathrm{r}})||^2/4,
\end{align}
where $(a_1)$ and $(a_2)$ hold if and only if $\mathbf{e}_{1,\mathrm{z}}^T(\bar{\mathbf{a}}_{\mathrm{t}}+\bar{\mathbf{a}}_{\mathrm{r}}) = 0$ and $\mathbf{e}_{1,\mathrm{z}}$ is parallel to $(-\bar{\mathbf{a}}_{\mathrm{t}}+\bar{\mathbf{a}}_{\mathrm{r}})$, respectively. Thus, when the $\bar F\left( \mathbf{\Omega}_1 \right)$ is maximized, we have $\mathbf{e}_{1,\mathrm{z}}^* = \frac{-\bar{\mathbf{a}}_{\mathrm{t}}+\bar{\mathbf{a}}_{\mathrm{r}}}{||-\bar{\mathbf{a}}_{\mathrm{t}}+\bar{\mathbf{a}}_{\mathrm{r}}||} = \frac{\mathbf{\Delta}_1}{||\mathbf{\Delta}_1||}$,
where $\mathbf{\Delta}_1 = (\Delta_{1,\mathrm{x}},\Delta_{1,\mathrm{y}},\Delta_{1,\mathrm{z}})^T$. It follows that $-(\mathbf{e}_{1,\mathrm{z}}^*)^T \bar{\mathbf{a}}_{\mathrm{t}} = (\mathbf{e}_{1,\mathrm{z}}^*)^T \bar{\mathbf{a}}_{\mathrm{r}} = \frac{1-\bar{\mathbf{a}}_{\mathrm{t}}^T\bar{\mathbf{a}}_{\mathrm{r}}}{||-\bar{\mathbf{a}}_{\mathrm{t}}+\bar{\mathbf{a}}_{\mathrm{r}}||} > 0$, unless $\mathbf{p}_{\mathrm{B},1}$, $\mathbf{p}_{1,1}$, and $\mathbf{p}_{2,1}$ are collinear. Without loss of generality, \eqref{Bat1} and \eqref{2at1} holds. Based on ${{\mathbf{e}}_{1,\mathrm{z}}} = {\left( {\cos {\theta _1}\cos {\varphi _1},\sin {\theta _1}\cos {\varphi _1},\sin {\varphi _1}} \right)^T}$, we have $\cos {\theta _1^*}\cos {\varphi _1^*} = \frac{\Delta_{1,\mathrm{x}}}{||\mathbf{\Delta}_1||}$, $\sin {\theta _1^*}\cos {\varphi _1^*} = \frac{\Delta_{1,\mathrm{y}}}{||\mathbf{\Delta}_1||}$, and $\sin {\varphi _1^*} = \frac{\Delta_{1,\mathrm{z}}}{||\mathbf{\Delta}_1||}$. The optimal azimuth and elevation angles are given by $\theta_1^* = \arctan \frac{\Delta_{1,\mathrm{y}}}{\Delta_{1,\mathrm{x}}}$ and $\varphi _1^*=\arcsin \frac{\Delta_{1,\mathrm{z}}}{||\mathbf{\Delta}_1||}$,
which completes the proof.

\bibliographystyle{IEEEtran}
\bibliography{refs.bib} 

\end{document}